\documentclass[conference]{IEEEtran}
\IEEEoverridecommandlockouts

\usepackage{cite}
\usepackage{amsmath,amssymb,amsfonts}
\usepackage{algorithmic}
\usepackage{graphicx}
\usepackage{textcomp}
\usepackage[dvipsnames]{xcolor} 
\usepackage[nolist]{acronym}
\usepackage{tabularx}
\usepackage{makecell} %
\usepackage{multirow}
\usepackage{listings}
\usepackage{tikz,pgfplots}
\usepackage{pgfplotstable}
\usepackage{xcolor}
\usepackage{fancyhdr}

\definecolor{cb_india}{HTML}{e5af53}
\definecolor{cb_cyan}{HTML}{0173b2}
\definecolor{cb_aquamarine}{HTML}{029e73}
\definecolor{cb_lilac}{HTML}{cc78bc}
\definecolor{cb_caramel}{HTML}{C91F37}
\definecolor{cb_dandelion}{HTML}{ece133}
\definecolor{cb_yellow}{HTML}{ede132}
\definecolor{cb_blue}{HTML}{0072b2}
\definecolor{cb_red}{HTML}{d55f01}

\definecolor{Charcoal Gray}{HTML}{1C1C1C}
\definecolor{Lavender Gray}{HTML}{6E75A4}
\definecolor{Prussian Blue}{HTML}{113285}
\definecolor{Tapestry Blue}{HTML}{0C4842}
\definecolor{Marine Blue}{HTML}{0D5661}
\definecolor{Midnight Blue}{HTML}{0B1013}
\definecolor{Smoke Blue}{HTML}{577C8A}
\definecolor{Marigold}{HTML}{FFB11B}
\definecolor{Raspberry Red}{HTML}{8E354A}
\definecolor{Tomato Red}{HTML}{F75C2F}
\definecolor{Malachite Green}{HTML}{227D51}

\pgfplotsset{
	compat = newest,
	tick label style={font=\sffamily\scriptsize},
	label style={font=\sffamily\scriptsize},
	legend style={font=\sffamily\scriptsize\raggedleft},
	legend cell align=left,
	grid style={dotted,gray}
}

\def\BibTeX{{\rm B\kern-.05em{\sc i\kern-.025em b}\kern-.08em
    T\kern-.1667em\lower.7ex\hbox{E}\kern-.125emX}}

\begin{document}

\title{X-WAD: eXplainable Web Anomaly Detection}
\author{
    \IEEEauthorblockN{Matteo Bitussi, Roberto Doriguzzi-Corin}
    \IEEEauthorblockA{\textit{Center for Cybersecurity, Fondazione Bruno Kessler, Italy}}
}

\maketitle
\thispagestyle{fancy}
\begin{acronym}[ModernBERT-large]
    \acro{srbh}[SR-BH2020]{SR-BH2020 dataset}
    \acro{srbhfix}[SR-BH2020-fix]{Fixed SR-BH2020 dataset}
    \acro{csicv2}[CSIC2010v2]{The CSIC 2010 dataset, in v2 version}
    \acro{csic}[CSIC2010]{The CSIC 2010 dataset}
    \acro{roberta}[RoBERTa]{Robustly Optimized BERT Pretraining Approach}
    \acro{robertahttp}[RoBERTa-http]{RoBERTa model pretrained on generic HTTP traffic}
    \acro{robertabase}[RoBERTa-base]{Smaller size RoBERTa model}
    \acro{smol}[SmolLM2-360M]{SmolLM2 model with 360 million parameters}
    \acro{modernbert}[ModernBERT-large]{ModernBERT model}
    \acro{bert}[BERT]{Bidirectional Encoder Representation Transformer}
    \acro{mlm}[MLM]{Masked Language Modeling}
    \acro{clm}[CLM]{Causal Language Modeling}
    \acro{llm}[LLM]{Large Language Model}
    \acro{tlm}[TLM]{Transformer-based Language Model}
    \acro{cnn}[CNN]{Convolutional Neural Network}
    \acro{rnn}[RNN]{Recurrent Neural Network}
    \acro{nn}[NN]{Neural Network}
    \acro{knn}[KNN]{K-Nearest Neighbors}
    \acro{svm}[SVM]{Support Vector Machine}
    \acro{gan}[GAN]{Generative Adversarial Network}
    \acro{lstm}[LSTM]{Long Short-Term Memory}
    \acro{plm}[PLM]{Pretrained Language Model}
    \acro{auc}[AUC]{Area Under Curve}
    \acro{aucpr}[AUC-PR]{Area Under Curve Precision-Recall}
    \acro{pr}[PR]{Precision-Recall Curve}
    \acro{shap}[SHAP]{SHapley Additive exPlanations}
    \acro{bbpe}[BPE]{Byte-level Byte Pair Encoding tokenizer}
    \acro{bpe}[BPE]{Byte Pair Encoding tokenizer}
    \acro{waf}[WAF]{Web Application Firewall}
    \acro{cot}[CoT]{Chain-of-Thought}
    \acro{roc}[ROC]{Receiver Operating Characteristic}
    \acro{tp}[TP]{True Positive}
    \acro{tn}[TN]{True Negative}
    \acro{fp}[FP]{False Positive}
    \acro{fn}[FN]{False Negative}
    \acro{tpr}[TPR]{True Positive Rate}
    \acro{fpr}[FPR]{False Positive Rate}
    \acro{fnr}[FNR]{False Negative Rate}
    \acro{tnr}[TNR]{True Negative Rate}
    \acro{apt}[APT]{Advanced Persistent Threats}
    \acro{ppl}[PPL]{Perplexity}
    \acro{capec}[CAPEC]{Common Attack Pattern Enumeration and Classification}
    \acro{xss}[XSS]{Cross-Site Scripting}
    \acro{cdn}[CDN]{Content Delivery Network}
    \acro{nlp}[NLP]{Natural Language Processing}
    \acro{ourtool}[X-WAD]{eXplainable Web Anomaly Detection}
    \acro{xai}[XAI]{Explainable Artificial Intelligence}
    \acro{pdf}[PDF]{Probability Density Function}
\end{acronym}

\begin{abstract}
The rapid growth of web-based services, particularly API-driven architectures, reflects an increasing reliance on distributed systems, exposing sensitive data to security risks and making the adoption of automated defensive mechanisms essential.
In this context, where benign traffic predominates in real-world settings, modern defenses increasingly model normal behavior, relying on semi-supervised approaches trained on only normal data. However, ensuring the complete absence of anomalous instances in such training data is inherently difficult in practice, and mislabeled or contaminated attack samples can introduce backdoors into the learned defense, causing the model to silently misclassify certain attack patterns as normal behavior.
This paper investigates the effectiveness of \acp{tlm} in detecting anomalies in HTTP requests, focusing on providing detailed explanations for the detected anomalies.
The study employs token-level logit-based surprisal mapping to provide both an anomaly score and a direct, detailed explanation via a heatmap-like highlighting. The effectiveness of the proposed explainability approach is demonstrated by the discovery of labeling inconsistencies in a popular public dataset, revealing how anomalous contamination in the training data had induced backdoor-like failures in the detection models.
\end{abstract}

\section{Introduction}\label{sec:intro}

Major \acp{cdn} and cloud providers report a consistent increase in web application traffic in their annual reports, alongside growing concerns about the associated security risks
\cite{akamai2024digital}.
Web security is challenging due to the complexity of modern attacks, the continuous emergence of zero-day vulnerabilities, and the large volume and heterogeneity of web traffic. In this regard, given the predominance of benign traffic in real-world settings, modern web anomaly detection approaches focus on modeling the normal behavior, typically through semi-supervised methods trained exclusively on benign data. While this paradigm is practical and effective, it relies on the assumption that training data is free from anomalous contamination, an assumption that is difficult to guarantee in practice. Even a small number of mislabeled or injected attack samples can bias the learned model, potentially introducing backdoor-like behaviors that cause specific attacks to be misclassified as normal.

Recent work has explored the use of \acfp{tlm} for anomaly detection in HTTP data. These approaches can be broadly categorized into four groups: feature extraction  \cite{doi/10.1145/3555776.3577663}, classification \cite{10.1109/ACCESS.2022.3185748}, prediction- or reconstruction-based methods \cite{2103.04475, tsai2025anollm}, and prompting \cite{10.1145/3643916.3644408}. Among these, prediction- and reconstruction-based approaches are particularly aligned with the normality modeling paradigm, as they learn the distribution of benign traffic and identify deviations through token-level probabilities.
Although \ac{tlm}-based methods have proven effective, their complexity limits interpretability, which is crucial for practical deployment in security contexts, where analysts must understand the rationale behind detections to effectively investigate and respond to misclassifications, including backdoor-like behaviors.

To improve interpretability, we propose \acsu{ourtool} (\acl{ourtool}), an explainability tool that leverages token-level scores derived from output logits of \ac{tlm}-based anomaly detectors to provide fine-grained explanations of anomalous inputs. \ac{ourtool} has been conceived to support security analysts in investigating flagged inputs by highlighting the specific parts of a request that contribute most to its classification, thereby improving transparency and trust in the detection outcomes.

In this work, we first assess the effectiveness of semi-supervised \acp{tlm} for anomaly detection in HTTP requests. We compare two architectural paradigms: unidirectional models (\ac{clm} with \ac{smol} \cite{smol}) for prediction-based detection, and bidirectional models (\ac{mlm} with \ac{modernbert} \cite{2412.13663}) for reconstruction-based detection. Evaluation is conducted on the \ac{srbh} \cite{sr_bh_2020} dataset. \ac{srbh} is one of the few available datasets specifically dedicated to HTTP attack detection on a single server. It is widely recognized in the literature and frequently employed as a benchmark across various studies \cite{2407.18445, 10.23919/CNSM59352.2023.10327888, 10.1109/ATC58710.2023.10318852, 10.1109/JSAC.2025.3560040, doi.org/10.1016/j.cose.2024.104127, 10.1007/978-3-031-74127-2_39}. 
While both paradigms achieve strong anomaly detection performance, their complexity limits interpretability, making it difficult to understand the rationale behind their decisions, particularly in cases of misclassification.\\
\ac{ourtool} overcomes this limitation by mapping output logits to token-level \textit{surprisal} scores, quantifying how unexpected each token is under the learned distribution of normal data. This enables a heatmap-like visualization that highlights the parts of a request contributing most to its classification.

This explainability mechanism enables deeper inspection of model behavior and reveals an important practical issue: contamination in training data. In particular, our analysis uncovers labeling inconsistencies in the \ac{srbh} dataset, showing how anomalous samples included in the training set can induce backdoor-like failures in detection models. We assess these inconsistencies and provide a corrected version of the dataset.
The list of identified inconsistencies in \ac{srbh} and the source code of \ac{ourtool} are publicly available to facilitate further research and ensure reproducibility \cite{ourtool}.

\section{Related Work}\label{sec:related_work}

Given the scale and complexity of \acp{tlm} and other deep learning models, interpreting their predictions remains a significant challenge. The task of understanding how a model reaches a specific decision for a given input is commonly referred to as \ac{xai}. Explainability plays a critical role in practical deployments: it enhances trust in model outputs, supports error analysis and model refinement, and facilitates the identification and mitigation of potential biases. Consequently, \ac{xai} has emerged as a highly active research area, encompassing a wide range of methodological approaches \cite{2407.19200}. The most prominent categories are summarized below.

\begin{itemize}
\item \textbf{Feature attribution methods}: These approaches adopt a black-box perspective by systematically perturbing the input—e.g., masking or removing tokens—and observing the resulting changes in model output. The magnitude of these changes is used to assign importance scores to input components. A widely used method is \ac{shap}, which leverages concepts from cooperative game theory to estimate feature contributions. While model-agnostic and broadly applicable to \acp{tlm}, these methods are often computationally intensive due to the large number of required evaluations. \cite{wang2024comparative, 10.1186/s40537-024-00905-w} Moreover, recent studies have raised concerns about the reliability of \ac{shap} explanations, particularly in the context of anomaly detection, where the model's sensitivity to input changes may not accurately reflect its internal reasoning \cite{2002.11097, 2302.08160}.

\item \textbf{Attention-based methods}: These techniques follow a white-box approach by exploiting the internal structure of transformer models. Specifically, they analyze attention weights, which quantify the interactions between tokens within a sequence. Visualization through attention heatmaps provides intuitive insights into token relationships during processing. However, it has been argued that attention weights do not always faithfully reflect the model's underlying decision-making process \cite{1902.10186}.

\item \textbf{Self-explaining methods}: These approaches generate natural language explanations to justify model predictions. A prominent example is \ac{cot} prompting, where the model produces step-by-step reasoning prior to the final output. Although such explanations are highly interpretable for humans, they may suffer from unfaithfulness, whereby the generated reasoning does not accurately correspond to the model's internal computations \cite{barez2025chain}.
\end{itemize}

In the context of explainable anomaly detection, a few recent works have explored the application of \acp{tlm} with a focus on \ac{xai}; all of the following works fall into the feature-attribution category.

\textit{Sec2vec} \cite{doi/10.1145/3555776.3577663} addresses anomaly detection in HTTP traffic by comparing \ac{roberta} with alternative vectorization techniques for feature extraction from textual request representations. A random forest classifier is then trained exclusively on normal samples. For explainability, the authors employ \ac{shap} to estimate token-level importance. 

\textit{HTTP2vec} \cite{2108.01763} systematically removes individual tokens, generates new embeddings for the altered request, and compares the resulting anomaly scores against the original to isolate each token's influence.
By removing tokens, there is a risk of creating malformed requests, causing the model trained on structurally correct data to potentially mislead the removal for a structural anomaly. Furthermore, removing only one token may leave a sufficient subset of the malicious syntax intact, keeping the anomaly score high. Consequently, the method underestimates the contribution of individual tokens that are part of a correlated structure, which may lead to incorrect feature importance scores.

\textit{ADALog} \cite{2505.13496} operates directly on unstructured log data, capturing intra-log contextual relationships and applying adaptive thresholding to normal samples. The method employs a transformer-based bidirectional encoder trained with \ac{mlm} and fine-tuned on benign logs. Anomalies are identified via token-level reconstruction probabilities, while interpretability is achieved through token-position analysis. Although its explainability approach is similar to that of this paper, ADALog proposes a dataset-level heatmap by aggregating patterns across many logs to show macroscopic trends in anomaly positions. However, this does not provide fine-grained explainability for a single sample to explain why it was flagged as anomalous.

\vspace{0.5cm}
Our method falls under the feature-attribution category; however, because we frame the anomaly detection task such that the output logits directly reflect model confidence, we can leverage these precalculated values for the input sample to obtain an explanation. Unlike \textit{HTTP2vec} or SHAP-based methods (\textit{Sec2Vec}), \ac{ourtool}'s approach incurs virtually no additional computational cost, as the output logits used for explainability are already computed when deriving the anomaly score.

\section{Threat Model}\label{sec:threat_model}
The threat model considered in this work assumes a web application exposed to the Internet, where adversaries interact with the system exclusively through HTTP requests. Attackers are modeled as external entities capable of crafting arbitrary request payloads to exploit input-validation weaknesses and application-logic flaws. The threat landscape includes a wide range of web-based attacks, such as injection attacks (e.g., SQL, command, and code injection), dictionary-based attacks, and path traversal. The model assumes that attacks are embedded within otherwise legitimate-looking traffic and may exhibit significant variability, including previously unseen (zero-day) patterns. Consequently, the detection approach must rely on learning the normal behavior of HTTP requests and identifying deviations from it, rather than depending on predefined signatures.

\begin{figure*}[t!]
  \centering
  \includegraphics[width=\textwidth]{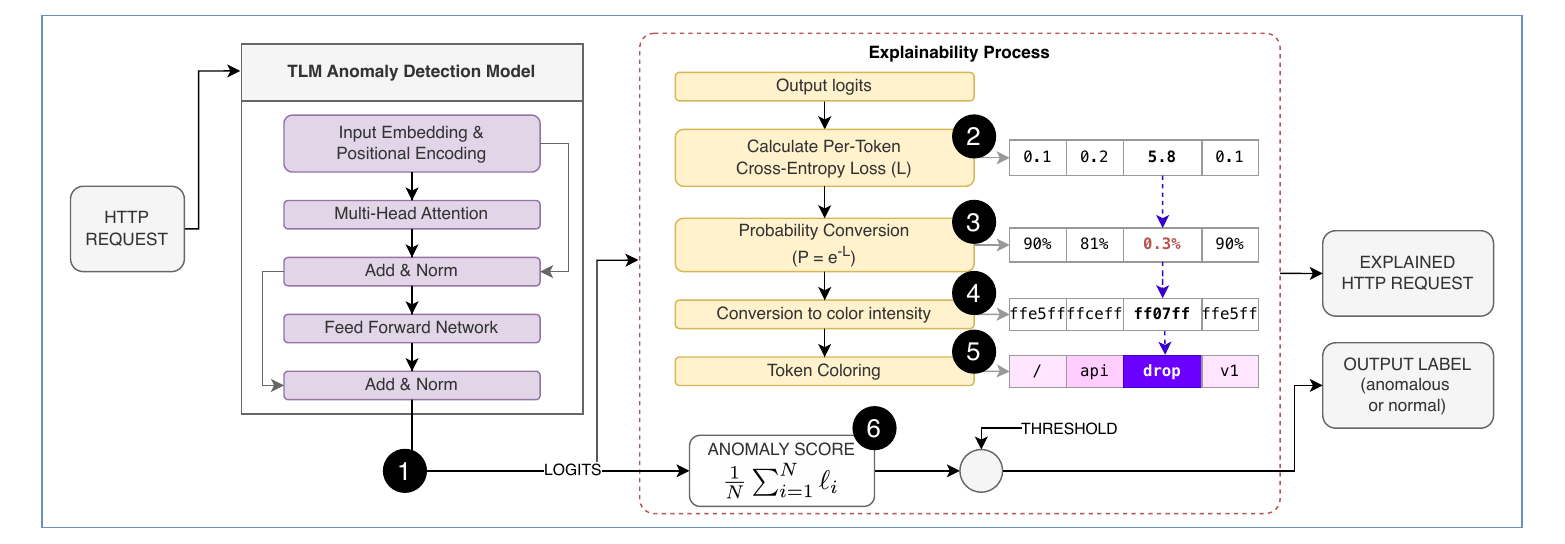}
  \caption{Inference pipeline.}
  \label{fig:inference_pipeline}
\end{figure*}
\section{Methodology}\label{sec:methodology}

The scale and high dimensionality of \acp{tlm} make their predictions inherently difficult to interpret. The methods implemented in \ac{ourtool} assign anomaly scores at the token level by highlighting segments of an HTTP request according to their contribution to the model's decision. Anomalous tokens are visually distinguished from normal ones through a color-encoded representation, enabling intuitive and fine-grained inspection of the model's behavior.
Notably, the proposed method is derived directly from the output logit distributions, thereby avoiding the need for additional inference steps that are usually required when using SHAP \cite{doi/10.1145/3555776.3577663} or other feature attribution methods \cite{2108.01763}.

The workflow of \ac{ourtool} is illustrated in Figure \ref{fig:inference_pipeline}. The diagram depicts the inference process, assuming a \textit{\ac{tlm} Anomaly Detection Model} that has been trained exclusively on benign data. The key idea is that the model learns the distribution of normal HTTP requests during training; consequently, any significant deviation from this learned distribution at inference time is flagged as anomalous. As represented in the figure, the output logits produced by the \ac{tlm} are directly leveraged to implement the \textit{Explainability Process}.

\subsection{Inference} 
The inference process varies depending on the model and its underlying task.

The \ac{clm} generates text in an autoregressive manner, meaning that each token is predicted based only on the tokens that precede it. This is implemented in a single forward pass using the same computations as during training, but without updating the model parameters. A causal (lower-triangular) attention mask ensures that each position can attend only to earlier tokens, enforcing the left-to-right dependency structure. Concretely, the model is given an input sequence and produces, for each position, a probability distribution over the next token. During training, the target sequence is defined as the input shifted by one position, so that each token prediction is aligned with its true next token. The prediction quality is measured at the token level using the cross-entropy between the predicted distributions and the corresponding target tokens.

In contrast, \ac{mlm} operates by masking one or more tokens within the input sequence and training the model to reconstruct them. Thanks to bidirectional attention, the model can use both left and right context to predict each masked token. At inference time, the process mirrors training: the model takes a tokenized input with masked positions and outputs probability distributions over the masked tokens, which can be compared to the true values using cross-entropy.\\
However, estimating token likelihoods is less straightforward than in autoregressive models. Since each token must be predicted without seeing itself, it needs to be masked at least once. To achieve this efficiently, a strided masking strategy is used: multiple inference passes are performed on the same input, each time masking a different subset of tokens.
The stride parameter $s$ controls how token masking is distributed across multiple inference passes. Instead of masking one token at a time (which would be computationally expensive), the sequence is processed $s$ times, each time masking a different subset of tokens spaced $s$ positions apart. In this way, multiple tokens are predicted in parallel during each pass, while ensuring that every token is masked, and therefore evaluated, exactly once across all passes. 
The value of $s$ determines the trade-off between efficiency and accuracy: smaller values allow more tokens to be processed per pass (faster computation), while larger values reduce interactions between masked tokens (more precise estimates). In this work, $s = 5$ is used as a practical compromise between speed and precision.

Regardless of the task used, the model ultimately produces output logits for each token (step (1) in the figure), which \ac{ourtool} uses to provide explanations by highlighting anomalous tokens and to calculate the anomaly score by averaging the per-token loss values.

\subsection{Explainability}
The per-token loss is calculated from the output logits using the cross-entropy loss function (step (2)). The model's assigned probability for each target token is then recovered using the relationship $P = e^{-L}$ (step (3)). The resulting per-token probability list is what is used to color the original sample tokens. A low predicted probability, correlated to a high cross-entropy loss, means that a specific token is unexpected within the given context based on the underlying (benign) data distribution learned during training. Therefore, the model interprets the presence of this token as an anomaly.
The visualization employs a color gradient to represent probability levels, ranging from lowest (purple) to highest (no color) (step (5)). The probability value of each token is mapped to a color intensity value that ranges from 0 to 255. This approach adapts to the specific probability distribution of each model and sample, highlighting rare events relative to their context.

\begin{figure}[th!]
  \centering
  \includegraphics[width=\linewidth]{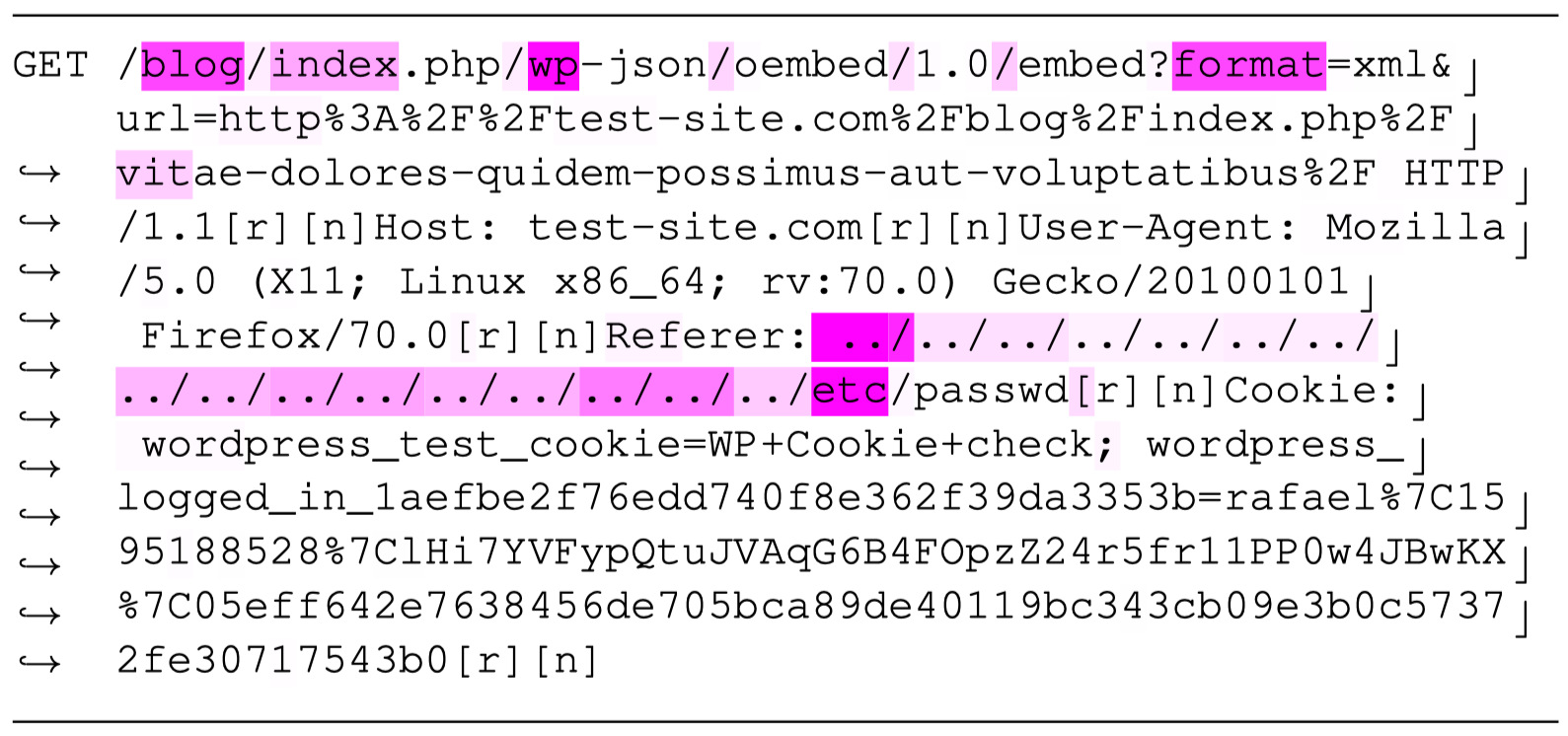}
  \caption{\ac{srbh} sample 27884 (anom.) explained with \acs{smol}.}
  \label{lst:srbh_sample_27884}
\end{figure}

An example of the result can be seen in Figure \ref{lst:srbh_sample_27884}, where an anomalous sample is highlighted according to the per-token probabilities.

The sample is classified as anomalous primarily due to the presence of the path traversal pattern \texttt{../../etc/passwd}. The anomalous nature of this observation is further supported by the pattern's location within the \texttt{Referer} header, representing a highly atypical placement. This visual explanation allows analysts to quickly identify the specific components of the request that contribute to its anomalous classification.

\subsection{Anomaly Score Calculation}\label{sec:methodology:score}
The anomaly score of a request is calculated using the model's output logits (step (6)). This is done by computing the average of the per-token loss values previously used for explainability:
$\frac{1}{N} \sum_{i=1}^{N} \ell_i$,
where $\ell_i$ is the loss value of the $i$-th token.
The resulting anomaly score is compared to a precomputed threshold: if it exceeds the threshold, the request is classified as anomalous; otherwise, it is considered benign.

\section{Experimental Setup}\label{sec:experimental_setup}

\subsection{Models}
The \acsp{tlm} employed in this work are \ac{modernbert} \cite{2412.13663} and \ac{smol} \cite{smol}, which utilize \ac{mlm} and \ac{clm} objectives, respectively, thereby covering both reconstructive and predictive paradigms. The two models have a comparable number of parameters (approximately 400M) and both employ a \ac{bbpe}.

\ac{modernbert} represents an architectural evolution of the original BERT framework \cite{1810.04805}. It retains the \ac{mlm} objective while extending the native context window to 8192 tokens. To reduce memory consumption, it incorporates rotary positional embeddings and adopts an alternating attention mechanism that combines global attention with local sliding-window attention. These design choices enable the model to achieve state-of-the-art performance while improving computational efficiency.
The model comprises approximately 395 million parameters, with 28 transformer layers, a hidden size of 1024, and 16 attention heads. The maximum supported context length is 8192 tokens.

\ac{smol} belongs to the second generation of the SmolLM series \cite{smol}. It is a decoder-only transformer architecture comprising 32 transformer layers and a hidden size of 960. The model is available in three configurations with 135M, 360M, and 1.7B parameters. In this work, we consider the 360M-parameter variant. The maximum context window supported by \ac{smol} is 2048 tokens.

\subsection{\acs{srbh} dataset}

The \ac{srbh} dataset from the work of Riera et al. \cite{sr_bh_2020} is a collection of HTTP requests gathered from a WordPress web server exposed to the internet. It is one of the few available datasets specifically dedicated to HTTP attack detection on a single server. It is widely recognized in the literature and frequently employed as a benchmark across various studies \cite{2407.18445, 10.23919/CNSM59352.2023.10327888, 10.1109/ATC58710.2023.10318852, 10.1109/JSAC.2025.3560040, doi.org/10.1016/j.cose.2024.104127, 10.1007/978-3-031-74127-2_39}.
The researchers manually and semi-automatically reviewed the tagging performed by a ModSecurity \ac{waf} that was installed alongside the web server, eventually correcting incorrectly labeled entries. The dataset is provided as a CSV file containing 525K normal and 382K anomalous HTTP requests, along with a binary label (benign/anomalous) and the \ac{capec} class for anomalous requests. A complete overview of the classes of the anomalous samples in the dataset can be found in Table \ref{tab:ano_class_distribution}.

\renewcommand{\arraystretch}{1.1} 

\begin{table*}[!t]

  \centering

  \caption{\acs{srbh} \acs{capec} class sample distribution.}

  \label{tab:ano_class_distribution}
  \footnotesize
  \begin{tabularx}{\textwidth}{>{\raggedright\arraybackslash}p{4.8cm} X c c}

    \Xhline{1.1pt}

    \textbf{CAPEC class} & \textbf{Description} & \textbf{Samples} & \textbf{\%} \\

    \Xhline{1.1pt}

    \textbf{Normal} & Benign HTTP requests & 525,193 & 57.85 \\ \hline

    \textbf{66 - SQL Injection} & Injection of crafted input into SQL queries to execute unintended database operations & 249,100 & 27.44 \\ \hline

    \textbf{194 - Fake the Source of Data} & The attacker spoofs request origin (e.g., via headers) to appear as a trusted source & 55,983 & 6.17 \\\hline

    \textbf{34 - HTTP Response Splitting} & Injection of control chars to split one HTTP response into multiple malicious responses & 19,668 & 2.17 \\\hline

    \textbf{126 - Path Traversal} & Access to restricted files by exploiting poor input validation (e.g., using "../") & 17,762 & 1.96 \\\hline

    \textbf{242 - Code Injection} & Injection of malicious code into the application & 15,805 & 1.74 \\\hline

    \textbf{272 - Protocol Manipulation} & The attacker manipulates the protocol to achieve unauthorized access & 9,153 & 1.01 \\\hline

    \textbf{88 - OS Command Injection} & Execution of malicious code on the server leveraging input validation flaws & 6,150 & 0.68 \\\hline

    \textbf{274 - HTTP Verb Tampering} & Modification of HTTP methods to bypass access controls & 4,055 & 0.45 \\\hline

    \textbf{310 - Scanning for Vulnerable SW} & The attacker probes the server to identify software versions and known vulnerabilities & 2,415 & 0.27 \\\hline

    \textbf{153 - Input Data Manipulation} & Alteration of input structure or format to exploit system weaknesses & 1,387 & 0.15 \\\hline

    \textbf{16 - Dictionary-based Password Attack} & Brute-force login attempts using common password lists & 1,142 & 0.13 \\

    \Xhline{1.1pt}
  \end{tabularx}
\end{table*}

As shown in the table, the dataset is unbalanced, with the majority of samples belonging to the normal class. The anomalous samples are distributed across 11 different classes, with varying frequencies. The dataset is also highly unbalanced in terms of the number of samples per class, with some classes having significantly more samples than others.

In the CSV file, each HTTP request is split into different components, such as the method, path, headers, and body (along with other less relevant fields). For the purposes of this work, we concatenate these components into a single string representing the entire HTTP request, which is then used as input to the models.

Although the dataset contains the response code and message generated by the web server for each HTTP request, we excluded this information from the training data to avoid introducing bias into the model.

\subsection{Training}

The training and validation sets are composed exclusively of normal samples, accounting for $70\%$ and $20\%$ of the normal data, respectively. The test set consists of the remaining $10\%$ of normal samples combined with all anomalous samples.
A batch size of $50$ is used for \ac{modernbert} and $12$ for \ac{smol}, with values chosen to maximize the utilization of the available VRAM on an NVIDIA RTX 3090 GPU. Due to memory constraints, gradient accumulation steps are employed to stabilize the training process. Early stopping is applied based on validation loss, halting training when performance on the validation set begins to degrade.

Additional implementation details differ across experiments but do not impact the explainability analysis. A complete specification of all training parameters is provided in the accompanying code repository \cite{ourtool}.

\subsection{Anomaly threshold tuning}
As described in Section \ref{sec:methodology:score}, samples whose anomaly score exceeds a threshold are classified as anomalous. The threshold $\tau$ is estimated using a Z-score approach, defined as
$\tau = \mu + k \cdot \sigma$,
where $\mu$ and $\sigma$ are the mean and standard deviation of the anomaly scores computed on the validation set, respectively, while $k$ is a tuning parameter used to calibrate the threshold according to the characteristics of the dataset.

\begin{figure}[h!]
  \centering

    \begin{minipage}[t]{\linewidth}
        \centering
    
        \input{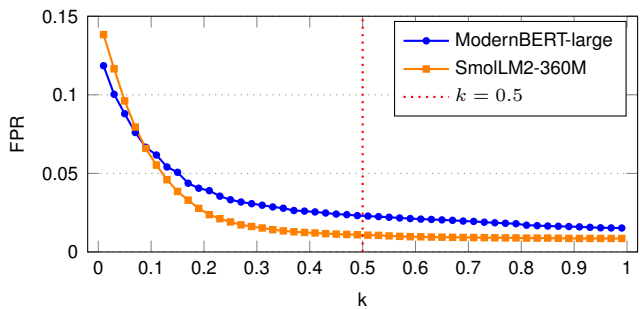}
        \pgfmathsetmacro{\lossmeanminus}{\lossmean - 0.5*\lossstd}
        \pgfmathsetmacro{\lossmeanplus}{\lossmean + 0.5*\lossstd}
    
        \pgfmathsetmacro{\lossmeanfour}{round(\lossmean*10000)/10000}
        \pgfmathsetmacro{\lossmeanminusfour}{round(\lossmeanminus*10000)/10000}
        \pgfmathsetmacro{\lossmeanplusfour}{round(\lossmeanplus*10000)/10000}
    
        \begin{tikzpicture}
          \begin{axis}[
              width=\linewidth,
              height=4.7cm,
              xlabel={k},
              ylabel={FPR},
              xmin=0, xmax=1,
              enlarge x limits=0.02,
              ymin=0, ymax=0.15,
              ymajorgrids=true,
              xtick={0, 0.1, 0.2, 0.3, 0.4, 0.5, 0.6, 0.7, 0.8, 0.9, 1},
              xticklabels={0, 0.1, 0.2, 0.3, 0.4, 0.5, 0.6, 0.7, 0.8, 0.9, 1},
              ytick={0, 0.05, 0.1, 0.15},
              yticklabels={0, 0.05, 0.1, 0.15},
              grid style={dotted,gray},
              tick label style={font=\sffamily\scriptsize},
              label style={font=\sffamily\scriptsize},
              legend style={draw=black, fill=white},
              legend plot pos=left
            ]
        
            \addplot[
              blue,                  
              thick,                 
              mark=*,                
              mark size=1pt,         
            ] table[
              col sep=comma,
              x=k,
              y=FPR
            ] {result-data/threshold_modernbert_srbh_fix.csv};
            \addlegendentry{\acs{modernbert}}

            \addplot[
              orange,                
              thick,                 
              mark=square*,                
              mark size=1pt,         
            ] table[
              col sep=comma,
              x=k,
              y=FPR
            ] {result-data/threshold_smol_srbh_fix.csv};
            \addlegendentry{\acs{smol}}
        
            \addplot[
              thick,
              dotted,
              red
            ] coordinates {
              (0.5,0)
              (0.5,1)
            };
            \addlegendentry{$k = 0.5$}
        
          \end{axis}
        \end{tikzpicture}
      \end{minipage}
    \caption{\acs{fpr} on the validation set at different values of $k$.}
    \label{fig:fpr_k_threshold}
\end{figure}

\begin{figure}[h]
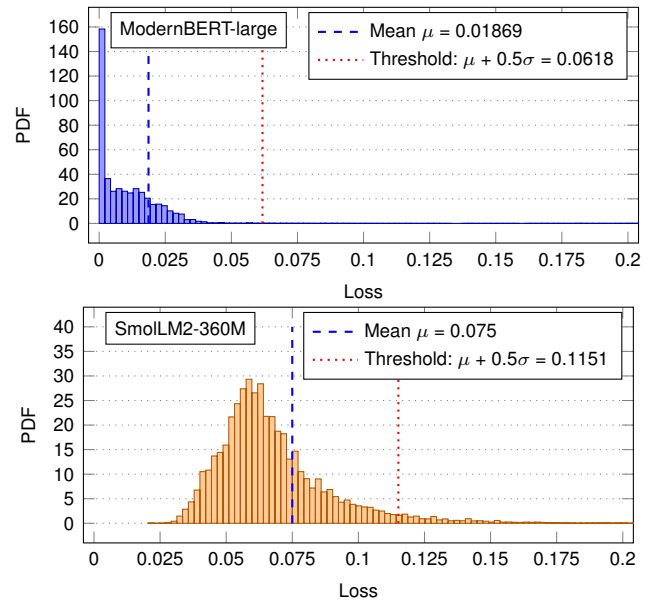

  \centering

  \begin{minipage}[t]{\linewidth}
    \centering

    \input{result-data/pred_srbh_val_modernbert-stats.tex}
    \pgfmathsetmacro{\lossmeanminus}{\lossmean - 0.5*\lossstd}
    \pgfmathsetmacro{\lossmeanplus}{\lossmean + 0.5*\lossstd}

    \pgfmathsetmacro{\lossmeanfour}{round(\lossmean*10000)/10000}
    \pgfmathsetmacro{\lossmeanminusfour}{round(\lossmeanminus*10000)/10000}
    \pgfmathsetmacro{\lossmeanplusfour}{round(\lossmeanplus*10000)/10000}

    \begin{tikzpicture}
      \begin{axis}[
          width=\linewidth,
          height=4.7cm,
          xlabel={Loss},
          ylabel={PDF},
          xmin=0, xmax=0.2,
          enlarge x limits=0.02,
          ymajorgrids=true,
          xtick={0, 0.025, 0.05, 0.075, 0.1, 0.125, 0.15, 0.175, 0.2},
          xticklabels={0, 0.025, 0.05, 0.075, 0.1, 0.125, 0.15, 0.175, 0.2},
          ytick={0, 20, 40, 60, 80, 100, 120, 140,160},
          yticklabels={0, 20, 40, 60, 80, 100, 120, 140,160},
          grid style={dotted,gray},
          tick label style={font=\sffamily\scriptsize},
          label style={font=\sffamily\scriptsize},
          legend style={draw=black, fill=white},
          legend plot pos=left
        ]

        \addplot[
          ybar interval,
          fill=blue!40,
          draw=blue!70!black,
          forget plot
        ] table[
          col sep=comma,
          x=bin_left,
          y=density
        ] {result-data/pred_srbh_val_modernbert-bins.csv};

        \addplot[
          thick,
          dashed,
          blue
        ] coordinates {
          (\lossmean,0)
          (\lossmean,160)
        };
        \addlegendentry{Mean $\mu$ = \lossmeanfour}

        \addplot[
          thick,
          dotted,
          red
        ] coordinates {
          (\lossmeanplus,0)
          (\lossmeanplus,160)
        };
        \addlegendentry{Threshold: $\mu$ + 0.5$\sigma$ = \lossmeanplusfour}

        \node[anchor=north west, fill=white, draw=black, font=\sffamily\scriptsize]
      at (rel axis cs:0.04,0.98) {\ac{modernbert}};

      \end{axis}
    \end{tikzpicture}
  \end{minipage}
  \hfill
  \\
  \begin{minipage}[t]{\linewidth}
    \centering

    \input{result-data/pred_srbh_val_smol-stats.tex}
    \pgfmathsetmacro{\lossmeanminus}{\lossmean - 0.5*\lossstd}
    \pgfmathsetmacro{\lossmeanplus}{\lossmean + 0.5*\lossstd}

    \pgfmathsetmacro{\lossmeanfour}{round(\lossmean*10000)/10000}
    \pgfmathsetmacro{\lossmeanminusfour}{round(\lossmeanminus*10000)/10000}
    \pgfmathsetmacro{\lossmeanplusfour}{round(\lossmeanplus*10000)/10000}

    \begin{tikzpicture}
      \begin{axis}[
          width=\linewidth,
          height=4.7cm,
          xlabel={Loss},
          ylabel={PDF},
          xmin=0, xmax=0.2,
          enlarge x limits=0.02,
          ymajorgrids=true,
          xtick={0, 0.025, 0.05, 0.075, 0.1, 0.125, 0.15, 0.175, 0.2},
          xticklabels={0, 0.025, 0.05, 0.075, 0.1, 0.125, 0.15, 0.175, 0.2},
          ytick={0, 5, 10, 15, 20, 25, 30,35,40},
          yticklabels={0, 5, 10, 15, 20, 25, 30,35,40},
          grid style={dotted,gray},
          tick label style={font=\sffamily\scriptsize},
          label style={font=\sffamily\scriptsize},
          legend style={draw=black, fill=white},
          legend plot pos=left
        ]

        \addplot[
          ybar interval,
          fill=orange!40,
          draw=orange!70!black,
          forget plot
        ] table[
          col sep=comma,
          x=bin_left,
          y=density
        ] {result-data/pred_srbh_val_smol-bins.csv};

        \addplot[
          thick,
          dashed,
          blue
        ] coordinates {
          (\lossmean,0)
          (\lossmean,40)
        };
        \addlegendentry{Mean $\mu$ = \lossmeanfour}

        \addplot[
          thick,
          dotted,
          red
        ] coordinates {
          (\lossmeanplus,0)
          (\lossmeanplus,40)
        };
        \addlegendentry{Threshold: $\mu$ + 0.5$\sigma$ = \lossmeanplusfour}

        \node[anchor=north west, fill=white, draw=black, font=\sffamily\scriptsize]
      at (rel axis cs:0.04,0.98) {\ac{smol}};

      \end{axis}
    \end{tikzpicture}
  \end{minipage}

  \caption{PDFs of the loss values on the \textbf{original} Validation Set.}
  \label{fig:pdf_two_models}
\end{figure}

The optimal value of $k$ is determined by evaluating the \ac{fpr} on the validation set (only benign samples) over a range of candidate values (Figure~\ref{fig:fpr_k_threshold}). Experimental results show that $k = 0.5$ corresponds to the point at which the \ac{fpr} begins to stabilize, providing a balanced operating point between sensitivity and robustness. In particular, larger values of $k$ would further reduce the \ac{fpr}, but at the cost of increasing the \ac{fnr} at runtime, potentially causing malicious samples to remain undetected.\\
Figure \ref{fig:pdf_two_models} shows the anomaly threshold applied to the \acs{pdf} obtained with the two models on the validation set.

\subsection{Evaluation}
The two \ac{tlm} models are evaluated on the test set of \ac{srbh} using the estimated anomaly detection threshold.
The evaluation is performed on unseen data, i.e., data not used during training or validation, to ensure an unbiased assessment of the models' generalization capabilities. The test set includes both benign (negative class) and anomalous samples (positive class), and is unbalanced toward the positive class, as it contains all the anomalous samples of the dataset (382K) and only a fraction of the benign samples (around 52K).

The model is evaluated using the following metrics:\\
\textbf{False Positive Rate (FPR) or False Alarm Rate}:
The proportion of benign samples that are incorrectly classified as malicious.\\
\textbf{False Negative Rate (FNR) or Miss Rate}:
The proportion of malicious samples that are incorrectly classified as benign.\\
\textbf{Recall (True Positive Rate (TPR) or Sensitivity)}:
The proportion of samples that have been correctly classified as malicious over all the samples that are actually malicious. TPR is useful when dealing with unbalanced datasets where the positive samples are fewer than the negative ones, as it focuses only on them.\\
\textbf{Precision (Positive Predictive Value (PPV))}:
The proportion of samples correctly classified as malicious among all samples classified as positive, including both correct and incorrect positive predictions.
Precision is particularly important in scenarios where the cost of false positives is high, as it measures the accuracy of the positive predictions made by the model. In the context of anomaly detection, a high precision indicates that when the model flags a request as anomalous, it is likely to be correct, which is crucial for minimizing false alarms and ensuring that security analysts can focus on genuine threats.\\
\textbf{F1 score}: 
F1 score is the harmonic mean of \textit{Precision} and \textit{Recall}. It can be expressed as:
$$F_1 = 2\cdot(\text{PPV} \cdot \text{TPR})/(\text{PPV} + \text{TPR})$$
The F1 score is a key metric when dealing with unbalanced datasets, as it provides a single measure that balances the trade-off between \textit{Precision} and \textit{Recall}. A high F1 score indicates that the model is effective at identifying true positives while minimizing false alarms.

\section{Results}\label{sec:results}
This section presents the experimental workflow adopted to evaluate \ac{ourtool} and analyze the impact of training data quality on anomaly detection performance. We begin by training and evaluating two \ac{tlm} models (\ac{smol} and \ac{modernbert}) on the original SR-BH2020 dataset, defining a statistically grounded decision threshold based on the distribution of validation losses, and highlighting limitations through test-time performance. We then leverage \ac{ourtool} to investigate model failures, revealing labeling inconsistencies and anomalous contamination in the training data.

\subsection{Model evaluation on the \acs{srbh} dataset}

The models are then evaluated on the test set using the thresholding procedure described in Section \ref{sec:experimental_setup}, and the resulting performance metrics are summarized in Table \ref{tab:results}. Both models show an unexpectedly high \ac{fnr}: 23.96\% for \ac{modernbert} and 14.82\% for \ac{smol}. Such values indicate that a portion of malicious requests remains undetected, undermining the practical reliability of both approaches. Extensive tuning of the training hyperparameters, including learning rates, batch sizes, and regularization strategies, did not yield meaningful reductions in the FNR.

\begin{table}[h!]
  \centering
  \renewcommand{\arraystretch}{1.2}
  \caption{Test set performance on the original dataset.}
  \begin{tabular}{l c c c }
    \Xhline{1.1pt}
    \textbf{Model}               & \textbf{F1}       & \textbf{FNR}  & \textbf{FPR}        \\ \Xhline{1.1pt}
    \ac{modernbert}    & 86.16\%  & 23.96\%    & 1.68\%    \\ \hline
    \ac{smol}           & 91.31\%  & 14.82\%    & 5.05\%    \\ \Xhline{1.1pt}
  \end{tabular}

  \label{tab:results}
\end{table}

\subsection{Explainability and dataset fixing}

To investigate the causes of the high misclassification rate, we applied \ac{ourtool} to analyze the false negative samples. The resulting explanations revealed that the models systematically failed to identify clear attack patterns as anomalous.

\begin{figure}[h!]
  \centering
  \includegraphics[width=\linewidth]{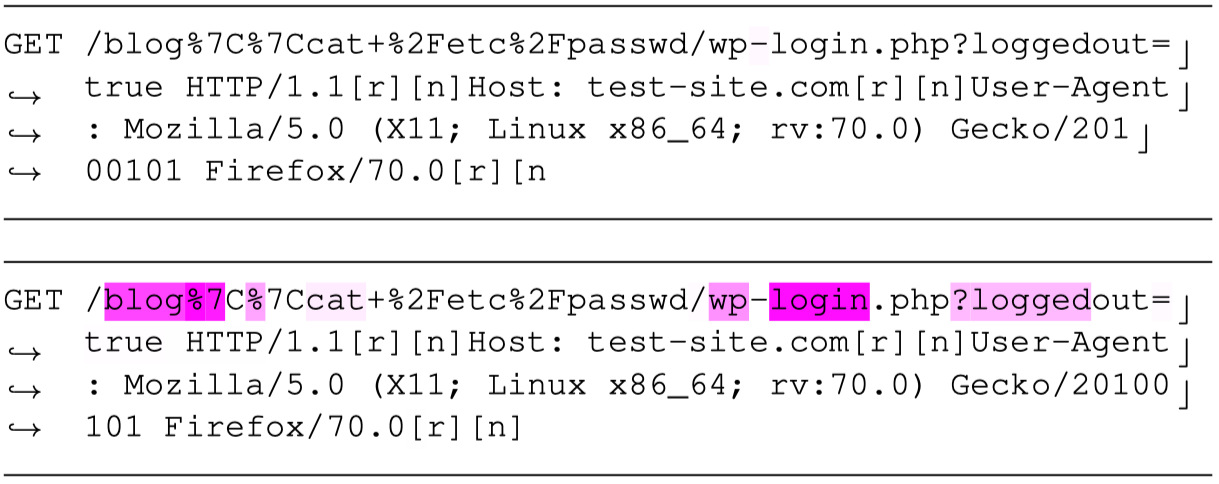}
  \caption{Sample 298021 explained by \ac{modernbert} (top) and \ac{smol} (bottom), both fine-tuned on \ac{srbh}.}
  \label{lst:srbh_sample_298021_original}
\end{figure}

A representative example is sample $298021$, whose explanations are reported in Figure \ref{lst:srbh_sample_298021_original}. Both models, particularly \ac{modernbert}, assign low (close to zero) anomaly scores to the substring \texttt{\%7C\%7Ccat+\%2Fetc\%2Fpasswd}, despite it being a well-known indicator of a command injection attack attempt, albeit in a malformed form given its position in the URI path.

This behavior suggests that such patterns have been incorporated into the learned notion of normality. A plausible explanation is that similar attack patterns are present in the training data labeled as benign. To verify this hypothesis, we searched for these patterns within the dataset and found that they appear not only in anomalous samples but also in a substantial number of samples labeled as normal. This observation indicates the presence of labeling inconsistencies.

Building on this insight, we systematically identified attack patterns from false negative samples using our explainability tool and searched for their occurrences within the subset of samples labeled as normal. This analysis revealed that approximately $13.8\%$ of the dataset is mislabeled, confirming that contamination in the training data significantly contributed to the observed degradation in detection performance.

Based on these findings, we created a revised version of the dataset with corrected labels, denoted as \ac{srbhfix}. The correction procedure involved reclassifying normal samples as anomalous if they exhibited the identified attack patterns.

To ensure the validity of this approach, a subset of the modified samples was manually inspected; however, due to the large scale of the dataset, exhaustive verification was not feasible. Consequently, some benign samples may have been incorrectly relabeled as anomalous. This may increase the number of false positives produced by the anomaly detector; however, unlike false negatives, which can introduce subtle backdoor-like behaviors in the model, such errors are explicitly exposed to the operator and can be further analyzed. These samples can be corrected and incorporated into subsequent training iterations, enabling iterative refinement of the model.

\begin{figure}[t!]
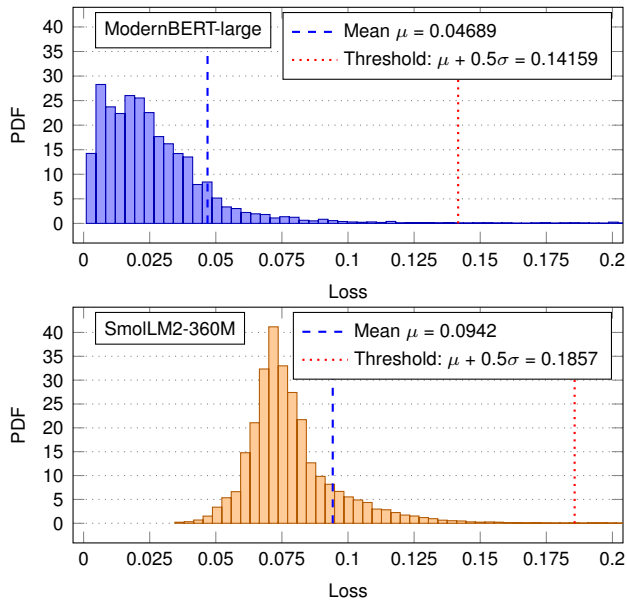

  \centering

  \begin{minipage}[t]{\linewidth}
    \centering

    \input{result-data/pred_srbhfix_val_modernbert-stats.tex}
    \pgfmathsetmacro{\lossmeanminus}{\lossmean - 0.5*\lossstd}
    \pgfmathsetmacro{\lossmeanplus}{\lossmean + 0.5*\lossstd}

    \pgfmathsetmacro{\lossmeanfour}{round(\lossmean*10000)/10000}
    \pgfmathsetmacro{\lossmeanminusfour}{round(\lossmeanminus*10000)/10000}
    \pgfmathsetmacro{\lossmeanplusfour}{round(\lossmeanplus*10000)/10000}

    \begin{tikzpicture}
      \begin{axis}[
          width=\linewidth,
          height=4.7cm,
          xlabel={Loss},
          ylabel={PDF},
          xmin=0, xmax=0.2,
          enlarge x limits=0.02,
          ymajorgrids=true,
          xtick={0, 0.025, 0.05, 0.075, 0.1, 0.125, 0.15, 0.175, 0.2},
          xticklabels={0, 0.025, 0.05, 0.075, 0.1, 0.125, 0.15, 0.175, 0.2},
          ytick={0, 5, 10, 15, 20, 25, 30,35,40},
          yticklabels={0, 5, 10, 15, 20, 25, 30,35,40},
          grid style={dotted,gray},
          tick label style={font=\sffamily\scriptsize},
          label style={font=\sffamily\scriptsize},
          legend style={draw=black, fill=white},
          legend plot pos=left
        ]

        \addplot[
          ybar interval,
          fill=blue!40,
          draw=blue!70!black,
          forget plot
        ] table[
          col sep=comma,
          x=bin_left,
          y=density
        ] {result-data/pred_srbhfix_val_modernbert-bins.csv};

        \addplot[
          thick,
          dashed,
          blue
        ] coordinates {
          (\lossmean,0)
          (\lossmean,40)
        };
        \addlegendentry{Mean $\mu$ = \lossmeanfour}

        \addplot[
          thick,
          dotted,
          red
        ] coordinates {
          (\lossmeanplus,0)
          (\lossmeanplus,40)
        };
        \addlegendentry{Threshold: $\mu$ + 0.5$\sigma$ = \lossmeanplusfour}

        \node[anchor=north west, fill=white, draw=black, font=\sffamily\scriptsize]
      at (rel axis cs:0.04,0.98) {\ac{modernbert}};

      \end{axis}
    \end{tikzpicture}
  \end{minipage}
  \hfill
  \\
  \begin{minipage}[t]{\linewidth}
    \centering

    \input{result-data/pred_srbhfix_val_smol-stats.tex}
    \pgfmathsetmacro{\lossmeanminus}{\lossmean - 0.5*\lossstd}
    \pgfmathsetmacro{\lossmeanplus}{\lossmean + 0.5*\lossstd}

    \pgfmathsetmacro{\lossmeanfour}{round(\lossmean*10000)/10000}
    \pgfmathsetmacro{\lossmeanminusfour}{round(\lossmeanminus*10000)/10000}
    \pgfmathsetmacro{\lossmeanplusfour}{round(\lossmeanplus*10000)/10000}

    \begin{tikzpicture}
      \begin{axis}[
          width=\linewidth,
          height=4.7cm,
          xlabel={Loss},
          ylabel={PDF},
          xmin=0, xmax=0.2,
          enlarge x limits=0.02,
          ymajorgrids=true,
          xtick={0, 0.025, 0.05, 0.075, 0.1, 0.125, 0.15, 0.175, 0.2},
          xticklabels={0, 0.025, 0.05, 0.075, 0.1, 0.125, 0.15, 0.175, 0.2},
          ytick={0, 5, 10, 15, 20, 25, 30,35,40},
          yticklabels={0, 5, 10, 15, 20, 25, 30,35,40},
          grid style={dotted,gray},
          tick label style={font=\sffamily\scriptsize},
          label style={font=\sffamily\scriptsize},
          legend style={draw=black, fill=white},
          legend plot pos=left
        ]

        \addplot[
          ybar interval,
          fill=orange!40,
          draw=orange!70!black,
          forget plot
        ] table[
          col sep=comma,
          x=bin_left,
          y=density
        ] {result-data/pred_srbhfix_val_smol-bins.csv};

        \addplot[
          thick,
          dashed,
          blue
        ] coordinates {
          (\lossmean,0)
          (\lossmean,40)
        };
        \addlegendentry{Mean $\mu$ = \lossmeanfour}

        \addplot[
          thick,
          dotted,
          red
        ] coordinates {
          (\lossmeanplus,0)
          (\lossmeanplus,40)
        };
        \addlegendentry{Threshold: $\mu$ + 0.5$\sigma$ = \lossmeanplusfour}

        \node[anchor=north west, fill=white, draw=black, font=\sffamily\scriptsize]
      at (rel axis cs:0.04,0.98) {\ac{smol}};

      \end{axis}
    \end{tikzpicture}
  \end{minipage}

  \caption{PDFs of the loss values on the \textbf{fixed} Validation Set.}
  \label{fig:pdf_two_models_fix}
\end{figure}

The resulting \ac{srbhfix} dataset was used to retrain the two models using the same method described above. The distribution of the predictions on the validation set from the fixed dataset is shown in Figure \ref{fig:pdf_two_models_fix}. The distribution of prediction losses on the normal validation set reveals a clear contrast in model behavior before and after dataset correction, particularly for \ac{modernbert}, where the model trained on the mislabeled dataset had a long convergence time ($18$ hours) and produced a PDF with an anomalous concentration of near-zero losses. This behavior suggests that the model overfit to the training distribution, memorizing specific data points to overcome the contradictory signals introduced by mislabeling. In contrast, training on \ac{srbhfix} required only $4$ hours and $30$ minutes, resulting in a loss PDF characterized by a wider, more continuous spread.
This reduced training time prevents the model from overfitting to individual samples, allowing it to capture the underlying, generalizable patterns in the data. Consequently, the broader dispersion of losses in the corrected model more accurately reflects the natural variance expected within the normal class, confirming a shift from detrimental memorization to healthy generalization.

\subsection{Model evaluation on fixed dataset}

The results on the revised dataset are reported in Table \ref{tab:results-fix}. Both models show a substantial improvement in detection performance, as reflected by the increase in F1 score and the marked reduction in \ac{fnr}. In particular, the F1 score reaches $98.29\%$ for \ac{modernbert} and $98.84\%$ for \ac{smol}, while the \ac{fnr} decreases from around $24\%$ and $15\%$ on the original dataset to $3.02\%$ and $2.12\%$, respectively.

\begin{table}[t!]
  \centering
  \renewcommand{\arraystretch}{1.2}
  \caption{Test set performance on the revised dataset.}
  \begin{tabular}{l c c c }
    \Xhline{1.1pt}
    \textbf{Model}               & \textbf{F1}       & \textbf{FNR}  & \textbf{FPR}        \\\Xhline{1.1pt}
    \ac{modernbert}    & 98.29\%  & 3.02\%     & 2.29\%    \\ \hline
    \ac{smol}           & 98.84\%  & 2.12\%     & 1.10\%    \\ \Xhline{1.1pt}
  \end{tabular}

  \label{tab:results-fix}
\end{table}

Figure \ref{fig:two_fnr_plots} further illustrates the reduction in \ac{fnr} across $10$ of the $11$ \ac{capec} classes, namely those that exhibited a non-zero \ac{fnr} in the original dataset for both models. The comparison between the original and fixed datasets highlights consistent improvements across all categories. The only exception is the “Protocol Manipulation” class, whose residual \ac{fnr} suggests that some attack patterns within this class may not have been fully captured during the dataset refinement process.

\begin{figure}[h]
\centering

\begin{minipage}{\linewidth}
\centering

\begin{tikzpicture}
    \begin{axis}[
        ybar,
        width=1\linewidth,
        height=4.7cm,
        bar width=6pt,
        enlarge x limits=0.08,
        ylabel={FNR (\%)},
        xlabel={},
        symbolic x coords={
          272 - Protocol Manipulation,
          126 - Path Traversal,
          88 - OS Command Injection,
          274 - HTTP Verb Tampering,
          66 - SQL Injection,
          153 - Input Data Manipulation,
          34 - HTTP Response Splitting,
          242 - Code Injection,
          194 - Fake the Source of Data,
          16 - Dictionary-based Password Attack
        },
        xtick=\empty,           
        xticklabels=\empty,     
        ymin=0,
        ymax=100,
        ytick={0,10,20,30,40,50,60,70,80,90,100},
        yticklabels={0,10,20,30,40,50,60,70,80,90,100},
        enlarge y limits={lower, value=0.02},
        legend pos=north east,
        ymajorgrids=true,
        scaled y ticks=false,
        yticklabel style={
          /pgf/number format/fixed,
          /pgf/number format/precision=2
        },
    ]

      \addplot[fill=blue!40,draw=blue!70!black]
      table[x=group, y=Original, col sep=comma]
      {result-data/original_fix_fnr_comparison_modernbert.csv};

      \addplot[fill=orange!40,draw=orange!70!black]
      table[x=group, y=Fix, col sep=comma]
      {result-data/original_fix_fnr_comparison_modernbert.csv};

      \legend{Original, Fix}

      \node[anchor=north, fill=white, draw=black, font=\sffamily\scriptsize]
      at (rel axis cs:0.5,0.98) {\ac{modernbert}};

    \end{axis}
\end{tikzpicture}

\end{minipage}

\begin{minipage}{\linewidth}
\centering

\begin{tikzpicture}
    \begin{axis}[
        ybar,
        width=1\linewidth,
        height=4.7cm,
        bar width=6pt,
        enlarge x limits=0.08,
        ylabel={FNR (\%)},
        xlabel={CAPEC class},
        symbolic x coords={
          272 - Protocol Manipulation,
          126 - Path Traversal,
          88 - OS Command Injection,
          274 - HTTP Verb Tampering,
          66 - SQL Injection,
          153 - Input Data Manipulation,
          34 - HTTP Response Splitting,
          242 - Code Injection,
          194 - Fake the Source of Data,
          16 - Dictionary-based Password Attack
        },
        xtick=data,
        xticklabels={
          \shortstack{272},
          \shortstack{126},
          \shortstack{88},
          \shortstack{274},
          \shortstack{66},
          \shortstack{153},
          \shortstack{34},
          \shortstack{242},
          \shortstack{194},
          \shortstack{16}
        },
        ymin=0,
        ymax=100,
        ytick={0,10,20,30,40,50,60,70,80,90,100},
        yticklabels={0,10,20,30,40,50,60,70,80,90,100},
        enlarge y limits={lower, value=0.02},
        legend pos=north east,
        ymajorgrids=true,
        scaled y ticks=false,
        yticklabel style={
          /pgf/number format/fixed,
          /pgf/number format/precision=2
        },
    ]

      \addplot[fill=blue!40,draw=blue!70!black]
      table[x=group, y=Original, col sep=comma]
      {result-data/original_fix_fnr_comparison_smol.csv};

      \addplot[fill=orange!40,draw=orange!70!black]
      table[x=group, y=Fix, col sep=comma]
      {result-data/original_fix_fnr_comparison_smol.csv};

      \legend{Original, Fix}

      \node[anchor=north, fill=white, draw=black, font=\sffamily\scriptsize]
      at (rel axis cs:0.5,0.98) {\ac{smol}};

    \end{axis}
\end{tikzpicture}

\end{minipage}
\vspace{-1mm}
\caption{FNR on \ac{srbh} and \ac{srbhfix}.}
\label{fig:two_fnr_plots}
\end{figure}

\begin{figure}
  \centering
  \includegraphics[width=\linewidth]{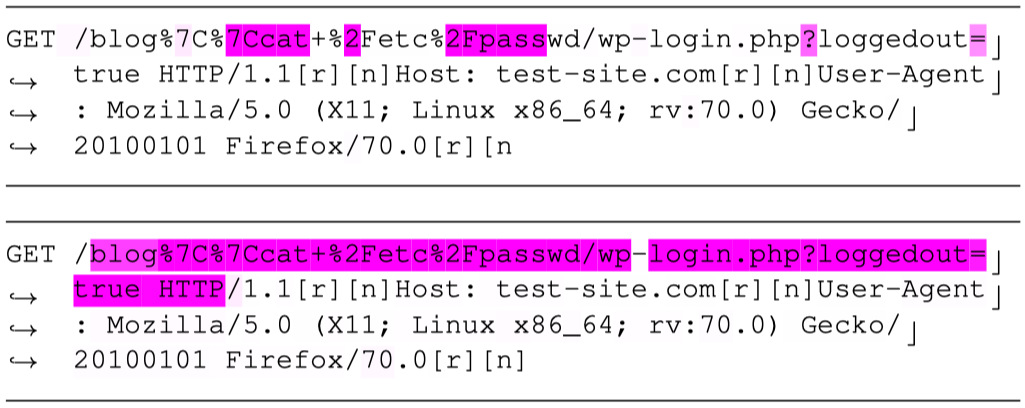}
  \caption{Sample 298021 explained by \ac{modernbert} (top) and \ac{smol} (bottom), both fine-tuned on \ac{srbhfix}.}
  \label{lst:srbh_sample_298021_fix}
\end{figure}

To confirm these findings, we reevaluate sample $298021$ using models trained on the \ac{srbhfix} dataset. The results are presented in Figure \ref{lst:srbh_sample_298021_fix}.

Both models correctly identify the \texttt{\%7C\%7Ccat+\%2Fetc\%2Fpasswd} pattern as anomalous, assigning high loss values to these tokens relative to their context. \\
While \ac{smol} highlights the entire anomalous sequence in solid purple, it is also more aggressive, flagging some benign tokens as anomalous. In contrast, \ac{modernbert} is more selective; although it leaves parts of the anomalous pattern unhighlighted, it produces significantly fewer false positives among normal tokens. 
These variations are partly driven by the different tokenization vocabularies, which determine the granularity and scope of the highlighted segments, and partly by differences in the underlying architectures of the models (masked and causal).

\section{Conclusions}\label{sec:conclusions}

The high \ac{fnr} observed on \ac{srbh} highlights the well-known sensitivity of anomaly detection approaches based on normality modeling to contamination in the training data. Given the scale of modern datasets and the inherent difficulty of exhaustive manual verification, it is crucial to assess whether a trained model captures the intended notion of normal behavior. If such contamination remains undetected in real-world deployments, it may introduce backdoor-like behaviors into the model, whereby specific attack patterns are systematically misclassified as benign and evade detection.

\ac{ourtool} enables the interpretation of model decisions by providing insight into why a given sample is classified as anomalous or benign, thereby allowing verification of whether the model behaves as intended. As demonstrated on \ac{srbh}, this facilitates the analysis of model outputs to assess whether well-known attack patterns are correctly identified as anomalous.

In our evaluation, the availability of a labeled test set allowed us to explicitly quantify performance metrics such as \ac{fnr} and \ac{fpr}, a condition that is not always met in real-world deployments. Nevertheless, we argue that similar validation can be achieved by applying the explainability method to randomly sampled benign-classified inputs during operation.

Although this work focused specifically on the explainability of HTTP requests, the proposed approach is generalizable to other forms of textual data. As future work, we plan to apply \ac{ourtool} to other data types, such as system logs or API calls, which are also commonly analyzed for anomaly detection. This would help verify the generalizability of the approach and potentially uncover similar issues in other datasets. Additionally, applying the tool in a real-world scenario would provide insights into its practical applicability and effectiveness in operational settings.

\section*{Acknowledgment}
This work was supported by Ministero delle Imprese e del Made in Italy (IPCEI Cloud DM 27 giugno 2022 - IPCEI-CL-0000007) and European Union (Next Generation EU).

\bibliographystyle{IEEEtran}
\bibliography{biblio}

\end{document}